# Memory in the system: trajectory-environment

Maxim Budaev*

1.08.2007

## Abstract

In this paper the memory effect in the system consisting from a trajectory of process and an environment is considered. The environment is presented by scalar potential and noise. The evolution of system is interpreted as process of the continuous "measurement" of a trajectory by an environment and\or on the contrary. In this sense the measured value is entropy-the information about embedding of the trajectory into configurations space. The concepts "measurement" and "interaction" are identified alike in quantum theory. Thus, the entropy evolution is not only the effect of dynamics, but also its cause, if it is possible to speak about causality in this case.

It is shown, that such non-elastic interaction leads to occurrence of memory in the system. The reflex memory mechanism is realized as a real conformal dynamic metrics. This is the drift factor, which is depending on all history of the process. In the paper the metastable event (attractor) is defined and time of its life is estimated.

It is shown, that in the system there is a change of dynamic modes from spatial localization to random walk (fluctuation, relaxation). On scale of the attractors' dynamics the memory is appearing as a semi-regular (in some information sense) drift of the image of fluctuations in a power spectrum of environment to absolute minimum. The latter properties (topological convergence) attract the attention to algorithmic aspect of a considered system.

______________________________________

*Email: budaev_mv@mail.ru

# 1. Introduction

The majority of the natural complex phenomena are non-stationary and non-equilibrium [1]. This circumstance complicates the application of a thermodynamic formalism for their analysis and modeling. Indeed, the non-stationarity does not assume presence ergodic (thermodynamic) limit and even denies its existence. Therefore there a potential infinity-length of a trajectory or "volume of habitat" in spatial sense is important. Thus communication between time and space because of absence ergodic limit becomes dynamic. This fact deduces on the advanced positions the individual trajectory of process as the basic object (and even a subject in our case as it will visible below) of the analysis. In our approach the space somewhat again under the attitude to time (trajectory) also serves only for the convenient indexation of the non-local events in time, or, in other words, to recurrencebility in what or sense.

In paper the multistable, stochastic system of gradient type is studied. Time is necessary discrete. The environment is presented by smooth scalar potential, stochastic noise and a partition of configuration space. In our model the interaction between a trajectory and an environment has the form of information interchange between them. Quantitatively the information is entropy; the entropy of a trajectory points distribution on some partition.

The partition, in our approach, is necessary for "measurement" of zero-dimension trajectory embedded in configurations space of an arbitrary dimension. We define a measure by means of partition. The partition can be interpreted also as the fact of the bounded measurement accuracy. As the part of an environment model, the partition compensates the smoothness of potential, replacing the convergence by stochasticity: in fact by virtue of the constructive assumption, an environment cannot be smooth. It is possible to tell also, that the partition is included in our model for weakening a continuum topology from a pre-ergodic (constructive) consideration.

In the considered system we include entropy in the dynamic equations as the "measured value", and we identify concepts "measurement" and "interaction". Thus, the entropy evolution is not only the effect of dynamics, but also its cause!

We shall show that such an inelastic dynamics leads to memory effect in the system evolution. For trajectory the information is reflexly present in the form of an "effective environment". The environment, in turn, "measures" a trajectory, absorbing the information on its spatial distribution.

The trajectory and the environment are representatives of dual categories, which interact by means of a real, conformal, dynamic metrics. This metrics non-monotomic depends on time as not increasing entropy function. We shall consider some interesting interpretations of this duality.

The Memory can be understood also as a phenomenon of presence of non-local time correlations or as non-local interactions. The correlations, as we shall see, can lead to effect of change of dynamic modes. Speaking by linear language of the quantum theory, it is a decoherence problem. The role of the memory in the mechanism of Anomalous Diffusion [1] is interesting. In paper the parallel with the theory hereditary processes [2] is considered.

We shall define the metastable event (attractor) and estimate the time of its life, as function of history of process and/or geometry of potential (landscape).

One of the memory demonstrations is existence of two is qualitative also quantitatively various dynamic modes:
- 1) "Fast" random walk on configurations space (relaxation). It is a spatially homogeneous phase.
- 2) Metastable events (fluctuations).

In the first case entropy increases, in the second one-decreases. We shall compare the times of a relaxation and fluctuation.

The important corollary of the memory in a considered system is a convergense on scale of set of attractors. It is possible to tell, that a spectrum of attractors parametrize the "global attractor".

The paper is organized as follows.
§2 is devoted to notations and terminology.
In §3 we denote an attractor (metastable event) and characterize him in terms convenient for following estimations. We prove the bifuracation relation (duality condition)–important result for local dynamic.
In §4 we estimate a lifetime of attractor and hence show what an attractor is not exactly an attractor; this object have some phases of evolution including a destruction phase. During the lifetime estimation we shall see these phases in detail. Here we denote an important value-the entropy production; the bifurcation production.
In §5 the scale of attractors dynamics is considered. It is shown that a system is degenerate into Marcovian trajectory on graph which is indexing the attractors. The sequence of lifetimes of attractors' is correlated by means of memory.
In §6 we shall consider the entropy evolution in more detail and estimate fluctuation time and relaxation time.
In §7 we illustrate information sense of a topological convergence.
In §8 various choice of partition as a part of environment model is considered.
§9 is devoted to discussion.

## 2. Process

In our model the memory of system is associated with metric parameter. The gradient flow of arbitrary potential can have a set of critical submanifolds of various dimensions-attractors, which draw the trajectories. Having added a stochastic term to gradient-like system it is possible to abstract from unstable, critical submanifolds. As a result the Langeven type system turns out:

$$\Delta X = -\lambda(t) \cdot \nabla_x U(X) \cdot \Delta t + \Delta \eta$$

$X \subset \mathbb{R}^n, n \in \mathbb{N}$ -configuration space,

$X(t)$-a culminating point of trajectory.

In future, for simplicity, we shall name $\{X(t)\}$ and $X(t)$ by "trajectory".

$U \in C^\infty(M)$-environment potential,

$\eta(t), \langle \eta(t) \rangle = 0$-stochastic process.              (1)

It is appropriate mention here about the nature of "noise". Actually it is the correlated noise which can not satisfy to the central limiting theorem. It is the non-stationary process having the nature of the considered trajectory (by virtue of duality, see below). This circumstance is not of great importance for our estimations and, in the further, we shall assume a noise is uniform distributed in a small bounded interval $(-\eta, \eta)$. The noise for us is important, first of all, as the "destruction mechanism" for unstable critical submanifolds of gradient flow.

$\lambda = \lambda(t) \in \mathbb{R}$ -is a bifurcation (renormalization) parameter-drift factor.

In fact it is the metric factor: $\lambda \equiv \lambda(t)\delta_{ij}$. For its definition we shall consider some partition $\xi$ of configuration space on events (coarse-grained). We shall put average diameter of a partition cell much more then noise amplitude- $\eta$. Define a dynamic frequency measure $\mu$ on elements of partition: Let by the moment $T$ the trajectory has spent in a cell $\xi_i$ of partition the time is equal $n_i(T)$. Then we define

$$\mu_i(T) \equiv \frac{n_i(T)}{T}.$$

Obviously,

$$\sum_{i \in V} \mu_i(T) = 1 \text{ for all } T.$$

Here $V(T) \subset \xi$ -set of "visited cells" by the moment $T$. We shall name this value by *volume*.

Thus, $\mu \in Sym(\mathbb{Q}^{|V|})$.

Now define $\lambda$.

**Definition 1**

The entropy of individual trajectory:

$$h(t) \equiv h(\xi, t) = -\sum_{i \in V} \mu_i(t) \cdot \log \mu_i(t),$$

Formally, it is Boltzmann's entropy of measures $\mu$ on a partition $\xi$.

Let

$$|V| \equiv |V(T)| \in \mathbb{N}$$ - is capacity of visited cells set. In the further we shall designate set of visited cells of partition and its capacity one symbol.

At last we shall put by definition:

$$\lambda(T) \equiv \lambda_0 \cdot \big(\log |V| - h(T)\big) \tag{2}$$

Here- $\lambda_0$ -positive constant, dimensional "the inverse mass".

The bifurcational parameter as not increasing function of entropy is a Lyapunov functional. The value of $\lambda$ can be considered also as the dynamic variable measuring a deviation of current distribution from uniform one. In this sense a bifurcational parameter is an information factor. On the other hand, it is scale of iteration (mobility) that is the power factor.

In the dual- $\lambda$ -functional is dynamic hierarchy of Riemannian conformal metrics, connected by the equation (2) which depends on the trajectory "mass distribution" in configuration space. Thus, the density fluctuations are expand/compress the space. In this sense $\lambda$ is a geometry factor.

## 3. Bifurcational relation

The bifurcational relation (duality condition) is an inequality establishing communication between the information and local geometry (energy) of an environment potential during the moment of transition between attractors.

First of all, we characterize attractors in terms convenient for further estimations. For this purpose we shall define attraction area. First we consider the elementary attractor-a local minimum. By virtue of the assumption of smoothness, in some vicinity of a local minimum the potential of an environment with some accuracy can be presented as a quadratic form. After reduction to main axes (proper basis) we obtain:

$$U(X) = U(X_0) + \sum_{i=1}^{n} \omega_i \cdot (X_i - X_0)^2 + o((X - X_0)^2).$$

Here $\omega_i \equiv \omega_{ii} \geq 0$ -is a local environment curvature. The equality sign corresponds to possible degeneration (extinct) in some directions in case of not trivial local minimum (critical submanifold). Let's emphasize, that local minima of potential can have various codimension.

**Definition 1**

Connected, open set $O$ such, that:

$$O(X_0) \equiv \forall X \in M : \left| U(X) - U(X_0) - \sum_{i=1}^{n} \omega_i \cdot (X_i - X_0)^2 \right| < \eta_0^3,$$

we shall name the basin of attraction or the parabolic vicinity of connected locally-minimal, critical submanifold

This definition is natural, as we have unique spatial smallness scale- $\eta_0$ (noise amplitude).

Generalization does not cause difficulties:

Let $M' \subset M, \|\partial_{M'} U\| = 0$ -a critical submanifold (a local minimum).

Then $O(M') \equiv \bigcup_{X \in M'} O(X)$.

In the further for estimations we shall characterize areas of an attraction by scalar:

$$\varpi = \frac{1}{N} \sum_{i=1}^{N} \omega_i$$ -is "local average curvature" or "energy" of attractor.

In estimations we shall use partition, associated with set of all attractors $A$:

$$\forall a \in A, \exists \xi_a \in \xi : O(a) \subset \xi_a,$$

and $A \leftrightarrow \xi$ -is a biunique correspondence.

Let the energy of attractor, which contains in a cell $\xi_i$, is $\varpi_i$. We shall name set $\{\varpi_i\}$ - a power spectrum of an environment.

Thus we identify the partition and the set of attractors and characterize them by the average curvatures:

$$\{A_i\} \leftrightarrow \{\xi_i\} \leftrightarrow \{\varpi_i\}.$$

In the end of paper it will be shown, that these assumptions are insignificant. Further the symbol $\varpi_i$ will designate event, attractor or attractor energy (curvature), depending on a context.

**Proposition 1**

The bifurcation relation is:

$$\lambda \cdot \omega > 1.$$
(3)

**Proof:**

The qualitative picture of system behaviour in a vicinity of a local minimum is those: After a trajectory fall into basin of attraction at time

$T: X(T) \in O(X_0), X_0 \in A$ it quickly rolls down on a bottom of a hole:

$$X \xrightarrow{t} X_0,$$

$$t \sim 1.$$

Then it is oscillate in a vicinity of a bottom with amplitude order of

$$|\Delta X| \simeq \lambda(T) \cdot 2 \cdot \varpi \cdot \eta_0.$$

Further, $\lambda$- function varies in the complicated manner until its value does not become sufficient for performance of a condition:

$$\lambda(T) \cdot 2 \cdot \varpi \cdot |\Delta X| > 2 \cdot |\Delta X|,$$

or

$$\lambda(T) > \varpi^{-1}.$$

Differently, the drift factor should change (increase) up to such magnitude that an each next iteration moves away a trajectory from "bottom" of a potential hole.

**Remark.**

The fact that $\lambda$-function grows in this case, will be discussed in detail at an estimation of attractor lifetime. It is also visible from qualitative picture: localization of a trajectory in attractor (measure concentration effect) is accompanied by entropy decreasing (increasing of conditional entropy) and by increase of $\lambda$-functions (2).

After performance of an inequality (3) trajectory it is very fast leaves the attractor. "Very fast" means:

$$\log \delta(t) \sim t^2,$$

$\delta(t)$-is a scale of a trajectory deviation from a bottom of potential hole.

Here $t$-a local time.

Really

$$\partial_t \delta \sim \lambda(t) \cdot \delta,$$

and, as it will be visible from the further(a curve into black rectangle on Figure 1), locally:

$$\partial_t \lambda \approx const > 0,$$

$$\lambda \sim t$$

We see, that a trajectory "blow-up tunneling" (superinflation) through a potential barrier. It states a necessary estimation and permits finish the proposition proof.•

The trajectory in space of potential variations

$$\varpi(t) \equiv \lambda_0^{-1} \cdot \lambda(t) \cdot \varpi \qquad (4)$$

is required to us. We shall name this by *effective (information) curvature* of potential. It is an illustration of a trajectory-environment duality.

Some important remarks it is appropriate mention here.

1. The blow-up inflation is allowing us to have no interest in the structure of a potential barrier (width, height, etc.). We shall assume that the trajectory have an "instant jump" into next attractor after performance the bifurcation relation (3).
2. If the trajectory gets in an attraction area and bifurcation relation (3) is attained, a attractor lifetime will be insignificant: $\tau \sim 1$. It is possible to tell, that the trajectory still "does not notice" of attractor. Or, in psychophysical terminology, the subject (trajectory) still remembers such or "more strong" geometry and does not the big attention (interest) for actual event. Take into account these reasons and that the individual trajectory instead of ensemble is considered, in the further we shall use concept of event instead of a state. Here pertinently to specify concept of metastability in the given context.

   **Definition 2**

   The event is called **metastable** or **singular** at performance of a condition:

   $$\varpi(t) \leq \lambda_0, \qquad (5)$$

   and **coherent** otherwise.

The metastable events quantitatively differ from coherent one greater time of life. The concept coherent is accepted in linear dynamics. In our approach it can be justified by following qualitative reasons: in case of metastable event the trajectory scans one attractor and in $\lambda$-functional the conditional information on attractor's geometry $\omega$ is accumulated. In a coherent case for arbitrary bounded "measurement time" (the time resolution is more than unit) in this measurement will take part a "superposition" of $n \approx \Delta t$ events-attractors. Differently, in the metastable case measurement is averaging on the outcomes belonging one event. In the coherent case averaging occurs on set of outcomes-events. In this definition it is important that the same event-attractor in due course can turn from metastable in coherent and \or on the contrary. Our technical identification attractor-event does not play a significant role here. This definition formalizes change of dynamic modes and will be discussed in detail in the further.

3. The bifurcation relation expresses the duality between an environment and a trajectory in terms $\varpi$ and $\mu$. This circumstance is reflection of information character of interaction between them.

4. In the further we shall not consider uninteresting, stationary situations: $\lambda_0 \cdot \log V < \varpi^{-1}$, is corresponding to delta-like density in a limit $t \to \infty$. From such stationarity it is possible to get rid, having used more fine partition or, speaking language of experimental physics, having increased the resolution of measurement tools.

## 4. Estimation of event's lifetime

**Definition 3**

Let at the moment $T$ the trajectory fall into basin of an attraction of some potential hole and has left them at the moment $T + \tau$. We shall name a value $\tau$ - time of a life (*lifetime*) of corresponding metastable event. In this sense, the frequency measure of attractor-event is:

$$\mu_i(T) = T^{-1} \sum_{r=1}^{R(T)} \tau_r ,$$

$R(T)$-a number of returnings of a trajectory in a cell $\xi_i$ in time $T$. This is relative, total life time of the event. It is important, that $\tau = \tau(T)$ time depended. This is illustration of time heterogeneity in our system.

For us it is necessary, by virtue of the offer (3) to estimate a time of evolution of bifurcation parameter to value $\varpi^{-1}$. We shall assume that attractor contains in a one cell of partition $\zeta_0$. Let's define small parameter:

$$\varepsilon = \frac{t}{T} \ll 1,$$

$T$-a time of an event $\zeta_0$ occurrence,

$t$-a current time of an event existence (6)

Thus, a global time is $T + t$. Here there are essentially various two cases:
1) The trajectory comes back in considered attractor.
2) The trajectory gets in "unknown attractor", increasing a volume in definition (2).

Let's consider these cases separately.

We shall name event $\xi_i$ actual, if $X(T+t) \in \xi_i$.

For small evolution it is had:

$$\mu(T+t) = \frac{n(T)+t}{T+t} = \frac{\mu(T)+\varepsilon}{1+\varepsilon}$$ -for actual event,

$$\mu(T+t) = \frac{n(T)}{T+t} = \frac{\mu(T)}{1+\varepsilon}$$ -otherwise.

Obtain:

$$\lambda(T+t) = \lambda_0 \cdot \left( \log V + \frac{\mu_0 + \varepsilon}{1+\varepsilon} \log\left(\frac{\mu_0+\varepsilon}{1+\varepsilon}\right) + \sum_{i=1}^{V-1} \frac{\mu_i}{1+\varepsilon} \log\left(\frac{\mu_i}{1+\varepsilon}\right) \right)$$

$$\lambda(T+t) > \varpi^{-1}.$$

It is easy to see, what the solving of last inequality is transcendentally complex. Therefore we shall receive square-law estimation. For this purpose we shall expand $\lambda$ in power series on small parameter $\varepsilon$ up to the second order. Necessity of the second order will be proved at calculations. It is also visible from Figure [1] and Figure [2].

We have:

$$\lambda(T+t) = \lambda(T, \varepsilon) = \lambda(T) + \lambda_\varepsilon \cdot \varepsilon + \lambda_{\varepsilon\varepsilon} \cdot \varepsilon^2 + o(\varepsilon^2).$$

After trivial calculations we obtain:

$$\lambda_\varepsilon = \lambda_0 \cdot (h(T) + \log \mu_0)$$

$$\lambda_{\varepsilon\varepsilon} = \lambda_\varepsilon + \frac{\lambda_0}{2} \cdot (\mu_0^{-1} - 1). \tag{7}$$

2) For iteration increasing volume $V \to V+1$ it is obtain:

$$\lambda(T+t) = \lambda(T) + \lambda_0 \cdot \log(1 + V^{-1}) + \lambda_\varepsilon \cdot \varepsilon + \lambda_{\varepsilon\varepsilon} \cdot \varepsilon^2 + o(\varepsilon^2).$$

Here:

$$\mu_0 \approx T^{-1},$$

$$\lambda_\varepsilon \approx \lambda_0 \cdot (h(T) - \log T),$$

$$\lambda_{\varepsilon\varepsilon} \approx \lambda_\varepsilon + \frac{\lambda_0}{2} \cdot T \text{, for greater } T.$$

In Figure [3] it is visible (smoothed), that after the occurense of unknown event there is the "stress"-splash of the information:

$\delta\lambda \sim \lambda_0 \cdot V^{-1}$ caused by topological properties of an environment. However, then the entropy very quickly increases:

$\partial_\varepsilon h \sim \log \varepsilon, \varepsilon \sim 0$, leading to a local relaxation Figure [3]. The stressful situation has two scenarios depending on a sign on value $\lambda + \delta\lambda - \varpi^{-1} = \lambda + \lambda_0 \cdot \log(1+V^{-1}) - \varpi^{-1}$ and "thickness" of walls of attractor. This or a panic flight (the red schedule), or a fright, replaced by intense interest (the green schedule) is well visible on Figure [3]. In the further we shall use the second scenario of behaviour for estimations. Here we shall not estimate accuracy of approach, being interested only order of magnitude and a qualitative picture.

We name the value $\lambda_\varepsilon$ by a conditional, normalized speed of change of entropy (metrics). Apparently, this value connects global $h$ and local $\log\mu$ information.
The sign of speed can be both positive, and negative. It explains two scenarios of behavior of bifurcation parameter, represented on Figure 1 and Figure 2.
For lifetime we obtain a rough estimate in linear approach:

$$\tau > \frac{\varpi^{-1} - \lambda(T)}{\lambda_\varepsilon} \cdot T = \frac{\varpi^{-1} - \lambda(T)}{\lambda_0 \cdot (h(T) + \log\mu_0)} \cdot T. \qquad (8)$$

For square-law approach it is necessary to find the solution of an inequality:

$$\lambda(T) + \lambda_\varepsilon \cdot \varepsilon + \lambda_{\varepsilon\varepsilon} \cdot \varepsilon^2 > \varpi^{-1}.$$

We obtain:

$$\tau > \frac{-\lambda_\varepsilon + \sqrt{\lambda_\varepsilon^2 - 4\lambda_{\varepsilon\varepsilon}\left(\lambda - \varpi^{-1}\right)}}{2\lambda_{\varepsilon\varepsilon}} \cdot T. \qquad (9)$$

Let's consider now two consecutive bifurcations: first the trajectory fall into attractor $\varpi_1$, then, after a while (lifetime), having tested bifurcation, in attractor $\varpi_2$ ($\varpi_1 > \varpi_2$)-differently the trajectory will not observe $\varpi_2$ by virtue of bifurcation relation (3). As it was already spoken it is possible to neglect the time of transition between attractors by virtue of it smallness in comparison with lifetime. We shall accept also $h > \log\mu_2$, as it is possible to use linear approximation for $\tau$, then:

$$\tau_2 \geq \left(\frac{1}{\varpi_2} - \frac{1}{\varpi_1}\right) \cdot \frac{T}{\lambda_0 \cdot \log V - \varpi_1^{-1} + \lambda_0 \cdot \log\mu_2}. \qquad (10)$$

These estimations look like an uncertainty relation (non-standard, renormalized). It is possible to interpret $\varpi$ as the factor of energy (the factor, not dependent on time). Really, locally in time (space) for eigenvalue of the operator of evolution it is obtain:

$$\partial_t X \approx E \cdot X,$$
$$E = E(t) = \lambda(t) \cdot \varpi.$$

For small time intervals, $\lambda \approx const$ and $E \sim \varpi$.
It justifies our interpretation. Thus, lifetime can be characterized in terms of the "energy jump":

$$\tau = \tau(\varpi_i \to \varpi_j) = \tau(\varpi_i, \varpi_j) = \tau_{ij}.$$

## 5. Factor system, "slow time"

Above, a local picture of the trajectory-environment interaction was considered. Now, we shall pass to more global point of view.
Let's consider the environment as a set of connected attractors. For this purpose we define the quotient partition. It is easy to see that this is topological oriented graph:

$$G = (V, E).$$

The attractors are corresponding to vertices vertexes $V$, and to edges $E$-possible transitions between them. Each vertex is characterized by effective curvature (energy) of corresponding attractor:

$$W : V \to \{\varpi\}.$$

Here $\{\varpi\}$-a power spectrum of an environment potential- set of attractors energies (average curvatures).

Two vertices vertex are connected by the edge:
$$e_{ij} \in E \subset V \times V,$$
if the direct transition between corresponding attractors in the mechanism of bifurcation (3) is possible. The direction of an edge is defined by a direction of a trajectory transition. For each vertex $v \in V$ we shall designate set of outgoing edges as $out(v)$. To each edge from this set we shall establish a correspondence with a transition probability between the vertex-attractors:
$$p : E \to (0,1), \sum_{j \in out(i)} p_{ij} = 1.$$
This randomness is caused by noise of model-$\eta$. In future, for simplicity, we shall consider a variant of a lattice:
$$p_{ij} = p_{ji} = (2 \dim M)^{-1}, \text{ if } i, j \text{-are indexes of the neighbouring cells}$$
and
$$p_{ij} = 0 \text{-otherwise.}$$
Actually, a distribution of transitive probabilities is generally non-uniform and depends on local geometry of the environment potential graph (curvature tensor). For simplicity of the next estimations we abstract from this information and accept the spherically symmetric potential holes as attractors.

The events lifetimes (waiting time) give natural partition of time. This partition is not homogeneous, what is display non-stationarity of our system. We shall take the factor of this partition and name this factor-time $\Theta$ as "slow time". In the spatial partition accepted by us, it is the sequence, indexing transitions between events-attractors. Thus we obtain the Markovian process $\mathrm{T}$ on the graph $G$ i.e. the spatio-temporal factor.

**Definition 2**.
 We shall name a *factorsystem* the triplet:
$$F \equiv (G, P, \mathrm{T}) \qquad (11)$$

By analogy to the theory of renewal processes [1] $\mathrm{T}$ is an operating (subordinated) Markovian process.
Underlying representation of the evolution operator on the graph is:
$$\mathrm{T} \circ \Theta = \Theta + 1,$$
$$\mathrm{T} \circ T = T \cdot (1 + \varepsilon),$$
$$\mathrm{T} \circ \xi_{i_\Theta} = \xi_{i_{\Theta+1}} \text{ -random transition in the next vertex (attractor),}$$
$$\mathrm{T}^* \circ \mu_i = \left(1 + \varepsilon \mu_i^{-1}\right) \cdot (1 + \varepsilon)^{-1} \cdot \mu_i \text{-for actual event (excitation vertex),}$$
$$\mathrm{T}^* \circ \mu_j = (1 + \varepsilon)^{-1} \cdot \mu_j \text{ - for irrelevant events,}$$
$$\mathrm{T}^* \circ \mu_{V+1} = \varepsilon \cdot (1 + \varepsilon)^{-1} \text{ - for "new event". } 1 \leq i, j \leq V.$$
$$\mathrm{T}^* \circ h(\mu) = h(\mathrm{T}^* \circ \mu),$$
$$\mathrm{T}^* \circ \lambda(h) = \lambda(\mathrm{T}^* \circ h) + \delta(\mu_{act}) \cdot \lambda_0 \cdot \log(1 + V^{-1}).$$

Here $\delta(\mu_{act})$-Dirac delta function equal to unit at occurrence of new event: $\mu_{act}(T) = 0$ (increase in volume) and to zero otherwise.

$\tau$ - lifetime of event. The estimation is given in (8).

$$\varepsilon = \frac{\tau}{T} \ll 1, T \text{ - time of event occurrence,}$$

We have the factor-dynamics:
$$\ldots \to \xi_{i_\Theta} \to \xi_{i_{\Theta+1}} \to \ldots \to T^n \circ \xi_{i_\Theta} \to \ldots \qquad (12)$$

and corresponding codynamics:
$$\ldots \to \varpi_{j_\Theta} \to \varpi_{j_{\Theta+1}} \to \ldots \to \varpi_{j_{\Theta+n}} \to \ldots \qquad (13)$$

in a power spectrum of an environment.

Bifurcation relation (3) gives quality to attractors (events). Therefore it is possible to define "dynamics of quality":
$$\ldots \to \theta_{k_\Theta} \to \theta_{k_{\Theta+1}} \to \ldots \to \theta_{k_{\Theta+n}} \to \ldots \qquad (14)$$

Symbol $\theta_k$ reflects metastable quality of attractor and has one of two values: metastable (singular), coherent.

## 6. Fluctuation, relaxation

It is almost obvious, that the $\lambda$-factor is not monotonous function of time. It leads to destruction and occurrence of the attractors (14). By way of illustration we shall estimate times of fluctuation and a relaxation and compare them.

Let's consider the following sequence:

$$h \xrightarrow{t_F} (h - \Delta h) \xrightarrow{t_R} h,$$

$t_F$ -time of fluctuation,

$t_R$ -time of relaxation. $\qquad (15)$

The entropy fluctuation corresponds to fluctuation of density in the bifurcation mechanism. Therefore, as $t_F$ it is possible to accept the estimation for lifetime of metastable event (8). It is evolution of conditional entropy. Provided that the trajectory is localized in some cell of partition (it is trapped by attractor, metastable event). This dynamic mode we shall name *the fluctuation mode*. The relaxation, in opposite, is an unconditional entropy growth. Such situation can arise, for example, in a compact case $V = const$ after next "minimal bifurcation" when the relation (3) is true and curvature-energy of attractor is minimal. A long time after bifurcation (relaxation time) a trajectory does not notice the others attractors and, thus, it is a sequence of coherent events. Thus, the system degenerates in an ordinary Markovian process on the graph vertex's which actually coincides with operating process (11). This dynamic mode we shall name *the relaxation mode*.

Let's consider a compact case $V = const$. We assume, that in time $t_R$ the trajectory visits each cell of partition on the average $t_R \cdot V^{-1}$-times. It corresponds to uniform, stationary distribution of operating process (11). This is that we name *pre-ergodicity consideration*.

In that case entropy as function of time has estimation:

$$h(T + t_F + t) \approx h(T + t_F) + h_\varepsilon \cdot \varepsilon,$$
$$h_\varepsilon = -(h + \overline{\log \mu}), \qquad (16)$$
$$\overline{\log \mu} = \frac{1}{V} \sum_i \log \mu_i (T + t_F).$$

This estimation for entropy production is analogously to (7).

$$\varepsilon = \frac{t_R}{T + t_F}. \qquad (17)$$

According to (15) it is necessary to estimate $\varepsilon_R$ in following equality:

$$\Delta h \approx -(h - \Delta h + \overline{\log \mu}) \cdot \varepsilon_R$$

Here

$$\Delta h \approx (h + \log \mu_0) \cdot \varepsilon_F \approx \lambda_0^{-1} \frac{\Delta \omega}{\omega^2},$$

that corresponds to entropy fluctuation at transition $\omega_1 \to \omega_2, \omega_1 > \omega_2$

$$\Delta \omega = \inf_{\omega' \in \{\omega\}} |\omega - \omega'| \text{-is the minimal spectral gap.}$$

In the first infinitesimal order we obtain

$$\frac{t_F}{t_R} \approx \frac{|\overline{\log \mu}| - h}{h + \log \mu_0} \approx \frac{\lambda_0 \omega^2 (|\overline{\log \mu}| - h)}{\Delta \omega}. \qquad (18)$$

For validity of linear approximation for entropy evolutions assume what: $h + \log \mu_0 > 0$.

It is interesting to compare this value to unit.

$$\frac{t_F}{t_R} \sim 1, |\overline{\log \mu}| + |\log \mu_0| \approx 2h$$

$$\frac{t_F}{t_R} > 1, |\overline{\log \mu}| + |\log \mu| > 2h \qquad (19)$$

Certainly, it is rough estimate. We see that ratio (18) is determined by the location of trajectory in power spectrum and by the power spectrum density. The dynamic mode of the fast diffusion exists in our system. It practically coincides with operating process (11). Figuratively speaking, such mode allows system to abstract from "the studied part of the graph" (configuration space) and to concentrate its "attention" and high transition properties: $\sqrt{D \cdot t} \sim \sqrt{(2 \cdot \dim M)^{-1} \cdot t_R}$ on search the low-energy attractors. From pre-ergodic reasons it is possible to replace, a subgraph: $\forall V' \in V : \varpi(V') > \varpi_0$ by one vertex with curvature (energy)-

$$\varpi' = |V'|^{-1} \cdot \sum_{i \in V'} \varpi_i > \varpi_0,$$ that is for coherent event in sense dynamic factorization. This situation we shall name *the second order dynamic factorization*.

This example illustrates presence of two dynamic modes in system. The first, fluctuation is characteristic localization of a trajectory in space of configurations. The second one is a relaxational–non local. These modes replace each other in such a manner that fluctuations drift in a power spectrum of an environment in a direction of a minimum of energy.

We come to the important consequence.

**Proposition-definition**

The Memory in system: trajectory-environment leads to asymmetry of time that generates the directed drift (in some information sense) of metastable events in a direction of reduction of energy:

$$\varpi_{i_1} > \varpi_{i_2} > .... \qquad (20)$$

**Proof.**

Let's consider metastable event $\varpi_i > \varpi_{\min}$. A next metastable event can have both greater and smaller energy. We shall estimate time $t^+$ of transition of a trajectory in metastable event: $\varpi_j < \varpi_i$ and time $t^-$ of transition in metastable event: $\varpi_k > \varpi_i$. Then we shall compare them.

Let $T$ - the moment of bifurcation,
$$\varepsilon^- = \frac{t^-}{T},$$
$$\Delta_+\varpi = \inf_{k:\varpi_k > \varpi_i}(\varpi_k - \varpi_i) \quad \text{the top spectral gap and let}$$
$\Delta_+\varpi \ll \varpi_i$.

Then in linear approach, using estimation for relaxation time (19), we obtain:

$$\varepsilon^- = \frac{t_R}{T} \approx \frac{\Delta_+\varpi}{\omega^2} \cdot \frac{1}{\lambda_0(h+\overline{\log \mu})} > 0$$

This is the lower bound estimation for return time $t^-$ of system in a power spectrum. It is easy to see, that for $t^+$ such limitation does not exist. As already next iteration can translate system in a metastable phase. For this purpose it is necessary in order among neighbours of actual attractor existed attractors with smaller energy (3).

More precisely, let
$$V^- \equiv \{\forall v \in V : \omega(v) < \varpi_i\}$$ -a subset of graph vertex's with energy smaller, than energy of actual attractor-vertex. Then the probability that a trajectory will reach metastable event with smaller energy during the relaxation time is:

$$p_{t_R}(\omega_i \to \omega_j, \omega_j < \omega_i) \sim \sum_{t \leq t_R}\left(1 - \frac{V^-}{V}\right)^{t-1} \cdot \frac{V^-}{V} > 0.$$

In this estimation the Bernoulli tests instead of the Markovian process are used, that is quite admissible at the most general assumptions of an environment. At the same time, as it was been shown, the probability of occurrence of metastable event with greater energy during this time ($t_R$) is strong equal to zero. Figuratively, in order an interest to the history has arisen, it is necessary to forget partially.

Here there is a phenomenon of topological convergence; the convergence in attractor's space (11) and in environment spectra. The term "topological convergence" we have in sense of dynamical factorization of factor-system. This is a dynamical non-monotonous reduction of partition to trivial one. The fact that the memory realized as the geometrical (metric) factor, leads to topological convergence is remarkable. It is possible to tell, that in our system the memory is a source of consciousness of the purpose.

The *topological convergense* is not conventional. Actually the system is nonstationary and an image of its evolution in a power spectrum of potential $\{\omega_i\}$ is not reduced to simple convergence (regular, weak etc.). More likely it reminds a turbulent flow. For illustration, even in case of degenerate global minimum ($\{\omega_{\min,i} > 0\}, 0 < i < n, n > 1$) a transitions between its components are possible. For this purpose, as it is easy to see, it is necessary that before the first passage time of a global minimum the condition

$$\lambda_0 \cdot \log V > \omega_{\min}^{-1} \quad \text{(3)} \text{ was satisfied.}$$

Continuing a parallel with quantum mechanics, it is possible to tell, that the system converges in "classical sense", as sequence of the decoherences (sequence of the "wave function collapses").

Even after a global minimum is achieved the interesting dynamics do not disappear. The Global minimum will collapse and large relaxation is raising a trajectory in power spectrum of environment. And scenario will repeat.

## 7. Operating process cycles. Interference

Here we shall quantitatively estimate the information sense in a phenomenon of the topological convergence. For this purpose we shall consider the cycles of operating process and their action in real "fast" time.

Let:
$$\{V_i\} = \ldots \to V_{i_\Theta} \to V_{i_{\Theta+1}} \to \ldots \text{ a trajectory of operating process,}$$

The corresponding sequence of events lifetimes:
$$\{\tau_i\} = \ldots \to \tau_{i_\Theta} \to \tau_{i_{\Theta+1}} \to \ldots$$

Unlike a parental process, $\tau_i$-is strongly correlated. This is illustration of the information memory.

The scheme of our example is following: we shall consider a cycle of operating process $\{V_i\}$ (11):
$$C(V_{i_0}, t_c) \equiv V_{i_0} \to V_{i_1} \to \ldots \to V_{i_N} \to V_{i_0}.$$

Let $t_c$ - length of a cycle (fast time)-
$$t_c = \sum_{k=0}^{N} \tau_{i_k}, \; \tau_{i_k} \text{ -attractors lifetimes.}$$

Any cycle is presented by the operator:
$$C^*(V_i, T, t_c) : \tau_i(T) \to \tau_i(T + t_c) \text{ (cocycle).}$$

The scenario can look so: first, the trajectory, at the moment $T$ is grasped by attractor $\varpi_{i_0}$, there lives in it some time (lifetime of metastable event $\tau_1(T)$), then leaves it and after time $t_c$ comes back again. For this time system can or only relax, or try the intermittent sequence of fluctuations and relaxations. We shall consider relaxation mode, as the most probable and simple. Clearly, that new event's lifetime $\tau_2 = \tau(T + \tau_1 + t_c)$, will differ from $\tau_1$ by virtue of heterogeneity of time.

Moreover,
$$\exists t_0 : \forall t_c < t_0, \tau_2(t_c) < \tau_1.$$

We shall estimate length of "minimal cycle", in this sense.

Figuratively being expressed, the trajectory is not stay too long there where was recently. This circumstance pays attention to the anomalous diffusion nature of a considered system.

We shall start realization of the plan. Let's use a linear approach for lifetime:
$$\Delta\lambda \approx \lambda_\varepsilon \cdot \varepsilon_c$$
$$\lambda_\varepsilon \approx \lambda_0(h + \overline{\log \mu}).$$

Here, as in (16)

$$\overline{\log \mu} \equiv |V|^{-1} \sum_i \log \mu_i .$$

For simplicity of estimation, we implicitly assume that a piece of the trajectory corresponding to "minimal cycle" will be uniformly distributed on all space of configurations. Obviously, it is not absolutely so. However, if we consider a cycle as a coherent (relaxation) phase, that is the most probable, our assumption is quite justified.

Obviously, $\lambda_\varepsilon \leq 0$.

We obtain for new, relative attractor's lifetime:

$$\varepsilon_2 = \frac{\tau_2}{T} \approx \frac{\Delta \lambda}{\lambda_0 (\mathrm{T}^{t_c} \circ h + \log \mathrm{T}^{t_c} \circ \mu_0)},$$

$$\mathrm{T}^{t_c} \circ h \approx h + h_\varepsilon \cdot \varepsilon_c = h - (h + \overline{\log \mu})\varepsilon_c,$$

$$\log \mathrm{T}^{t_c} \circ \mu_0 \approx \log \mu_0 (1 + \varepsilon_c)^{-1}.$$

We must estimate $\varepsilon_c$ in equality:

$$\varepsilon_1 = \varepsilon_2.$$

More accurately:

$$\varepsilon_2 = \varepsilon_1 (1 + \varepsilon_1 + \varepsilon_c).$$

Nevertheless, we interest the first order of smallness on $\varepsilon$. Therefore we shall consider the first variant.

Obtain:

$$\frac{-(h + \overline{\log \mu})\varepsilon_c}{h + (h + \overline{\log \mu})\varepsilon_c + \log \mu_0 - \log(1 + \varepsilon_c)} \approx \varepsilon_1,$$

or, in first order of smallness:

$$\frac{\tau_c}{\tau_1} \approx \frac{\varepsilon_c}{\varepsilon_1} \approx -\frac{h + \log \mu_0}{h + \overline{\log \mu}}.$$

Here it is possible to note, that for the bifurcation moment: $h = \log V - (\lambda_0 \omega_0)^{-1}$ (3). As one would expect obtained estimation does not differ from (18). We go to a conclusion, that time of correlation of the order of a lifetime. Their distinction is defined first of all by non-uniformity of distribution of points of a trajectory in configuration space. Quantitatively this heterogeneity can be expressed in terms $\log \mu_0, \overline{\log \mu}$.

We have obtained estimation for the period of the minimal cycle. Let's name this value by *the correlation time*.

As it is easy to notice, on set of the vertex-attractors visited in the past, operating process can be considered as sequence of the cycles closes. Such an "interference of fluctuations" can lead to dynamic factorization even the events with the least energy at present. Differently: frequent, fast returnings of a trajectory are destroy the metastable events, translating them into the coherent rank.

## 8. Choice of partition

In above estimations the partition, associated with attractors set was used. Here we shall show that this assumption does not influence a qualitative picture.

Really, if attractor is covered by the some number (>1) of partition cells, the fluctuation phase can be considered as a conditional relaxation (15); the relaxation inside an attractor. Let's consider some partition $\xi^1$.

Let attractor $A$ is covered by $N$ partition sells:

$$A = \{\xi_i\}, |A| = N,$$
$a \subset \mathbb{N} \mid \forall \xi_i \in A, i \in a$ - set of incidental cells indexes.

For entropy product it is had (17):

$$^1h_\varepsilon = -(h_1 + \overline{\log \mu}),$$

$$\overline{\log \mu} \overset{def}{=} |A|^{-1} \sum_{i \in a} \log \mu_i > h.$$

( 21)

Let's consider, for an example, other partition $\xi^2 < \xi^1$ differing $\xi^1$ themes, that attractor $A$ becomes covered by one cell:

$$\xi_a^2 : \mu(\xi_a^2) = \mu_a, \mu_a = \sum_{i \in a} \mu_i.$$

In this case:

$$^2h_\varepsilon = -(h_2 + \log \mu_a)$$

$$h_2 = h_1 - \mu_a \log \mu_a + \sum_{i \in a} \mu_i \log \mu_i$$

$$\Delta h_\varepsilon \equiv {}^2h_\varepsilon - {}^1h_\varepsilon = (\mu_a - 1) \log \mu_a - \sum_{i \in a} \mu_i \log \mu_i + \overline{\log \mu}$$

The trajectory distribution inside attractor can be assumed as uniform one:

$$\forall i \in a, \mu_i \approx |A|^{-1} \cdot \mu_a.$$

We obtain:

$$-\sum_{i \in a} \mu_i \log \mu_i + \overline{\log \mu} \approx (1 - \mu_a) \overline{\log \mu},$$

$$\Delta h_\varepsilon \approx (\mu_a - 1)(\log \mu_a - \overline{\log \mu}) \approx (\mu_a - 1) \log |A| < 0.$$

Apparently, the qualitative picture is kept. Moreover, the quantitative effects poorly depend on the partition fining.

## 9. Discussion

Let us summarize what we have achieved so far. We have considered dynamics of system on various scales and have obtained a following qualitative picture:
The fluctuations of trajectory density distribution are alternated with relaxations. During fluctuation the trajectory is localized in the some attractor and makes oscillation around of gradient flow equilibrium, absorbing the information on local geometry of an environment. Entropy, thus, eventually, decreases up to some threshold defined by a bifurcation condition (3). Further the trajectory "blow-up tunneling" into neighboring attractor. After a fluctuation or the sequence of fluctuations (soliton) the system is in relaxation phase.
Time of a relaxation depends on habitable volume, entropy and size of an actual spectral gap. In information representation it means, that a trajectory non-local and in the entropy production take part a set of coherent events. In this phase the trajectory actually becomes Marcovian (diffusion, entropy growth).
First of all we are interested in a fluctuation mode. After bifurcation the metrics contains information on an environment, namely, about local curvature of the potential graph or energy. It can be interpreted as a measurement process.
We were shown that the image of dynamics in a power spectrum of an environment regular in some information (information memory) sense. This large scale, topological convergence is an important, non-obvious property, that generated by a history dependent dynamics. Therefore, it is possible to consider a trajectory of our system as a search algorithm of attractor with the minimal energy (curvature).
Further I want to make conceptual comparison our system (TE) with other known multistable models.

The Quantum Mechanics (QM) have a two dynamics regime: the unitary evolution of states superposition (wave mechanics) and the non-unitary projection to one of states (wave function collapse, measurement) [3, 4, 6]. In TE the superposition is corresponds to the Markovian trajectory ("fast diffusion" on configuration space, time average vs. state average (superposition)) and the wave function collapse is corresponds to singular event into attractor. The attractor has a finite lifetime (non-constant) which depends on his curvature (formal energy) what corresponds to the uncertainty relation in QM. In contrast to QM, an attractor's lifetime also depend on the entire trajectory i.e. from the measurement history (history dependent).

The super-string theory has the enormous amount of vacuas (landscape) [8, 9, 10]. These vacuas have different properties such as dimension, topology, masses, coupling constants. In this situation there is a necessity to have a non-perturbation mechanism for exploring the landscape, may be in form of a dynamic system. It is very likely that the Inflationary multiverse together with the Anthropic Principle has given such opportunity. The Dynamics of selection is the eternal inflation.
In these remarkable models the absolute value of some scalar potential(vacua energy) is important. In contrast, TE characterize own vacuums in terms of local geometry of the potential i.e. a curvature tensor field vs. scalar field.
Here I want to make an ontology comparison TE with the Anthropic landscape paradigm and the Inflationary multiverse (ALIM) in more detail.

| TE | ALIM |
|---|---|
| Environment potential, configuration space | landscape |
| Bifurcation(sequence of bifurcation) is induced by the Memory based dynamics | Inflation (eternal inflation) |
| measure concentration | Quantum fluctuations |
| Entropy product after inflation-bifurcation (not constant) | Hubble constant |
| Curvature tensor field | Scalar potential |

There are important differences TE from ALIM. The first is the constructive (non actual) approach to concept of space (configuration space). And second one is "non causal" or the memory based interaction between attractors (vacuas). The last claim requires an explanation.

In TE cosmology the "observer" is included in a dynamic mechanism. I.e they is a subject of Multiverse in ALIM sense. Figuratively, the TE "observer" living in attractor (vacua) wittingly or not do a permanent measurement of the own theory "constants" as observed values of multiverse, step by step localized its values in a decreasing scope until the uncertainty of a "constants impetus" make "quantum fluctuation" to next metastable vacua with new constants(physics) etc.. New vacuum may be a "known" (not interesting), "forgotten" (recollected), "unknown" and a shade of these, according to theirs lifetime.

There is an intriguing application for the Memory models. This is a neuronet learning algorithm. In general, the neuronet [11] is a function space, which is possible to imagine as a weighed net or even a weighed graph with appropriate size and topology and non-linearity in vertexes. The Error Functional exists on this space. This functional may have a complex structure for the gradient flux, consisting from a set of attractors (vicinities of the minimal critical submanifolds) with different error values and separated by a domain walls. The goal is to find a global minimum. TE solves this problem after interpretation: a neuronet space as a TE configuration space and an error functional as a TE environment potential.

The Memory for a neuronet learning algorithm is not only the mechanism of improvement of convergence, but also is a possible formalization of the intuitively-clear psychophysical phenomenon.

The neuronet paradigm lay claim to formalize the Consciousness. It is difficult to imagine the Consciousness without the Memory.

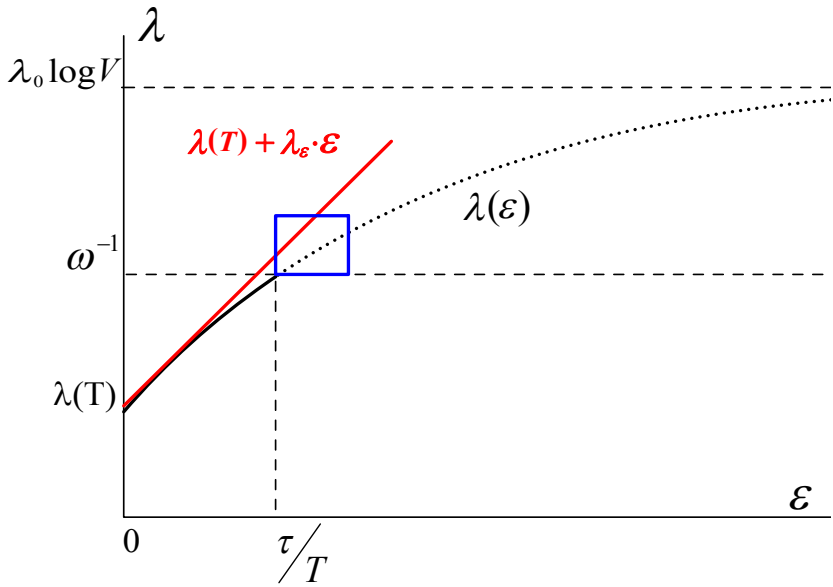

Figure 1: Linear approximation.

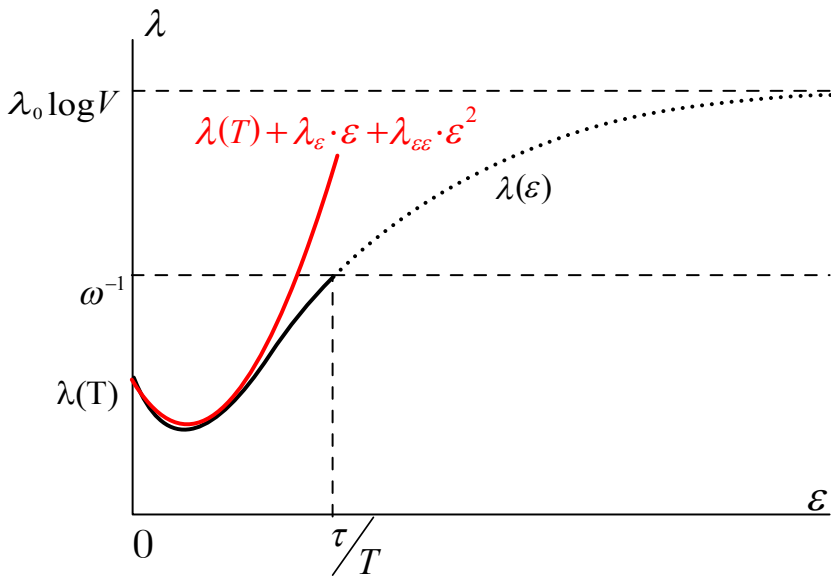

Figure 2: Quadratic approximation.

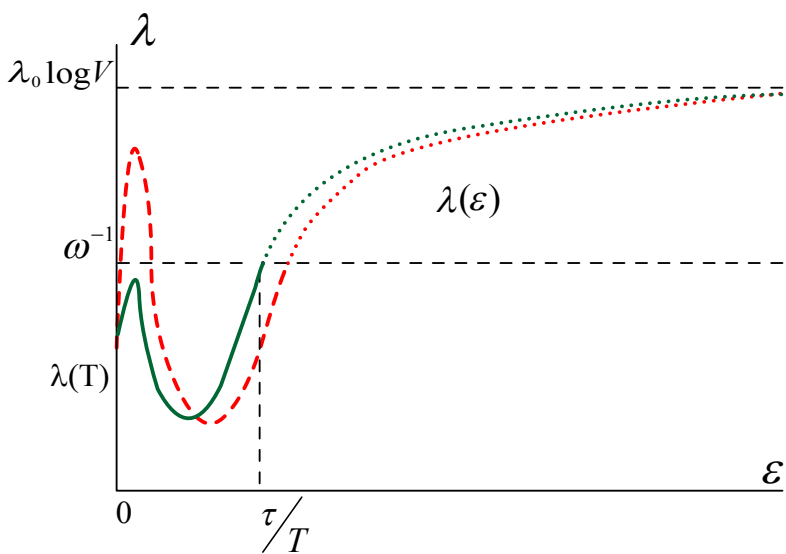

Figure 3: "Birth" of new event: $V \to V+1$, (smoothed).